\documentclass{elsartL}
\usepackage{graphicx}
\usepackage{amssymb}
\usepackage{amsmath}
\usepackage{physics}
\usepackage{mathrsfs}
\usepackage{mathtools}
\usepackage{rotating}
\usepackage{appendix}
\newcommand{\RNum}[1]{\uppercase\expandafter{\romannumeral #1\relax}}
\begin{document}

\begin{frontmatter}
\title{Analysis of four sets of global average temperature anomalies}
\author{Evangelos Matsinos}

\begin{abstract}
Four sets of global average temperature anomalies, altered so that they refer to pre-industrial temperature levels (baseline), as recommended by the Intergovernmental Panel on Climate Change, are 
analysed in this study. Expectation values of the time, at which the crossing of the $1.5^\circ$C threshold above the baseline will occur, are extracted from the two datasets which are accompanied by 
meaningful uncertainties; the two estimates are in the 2030s and differ by about four years: mid-2032 and mid-2036.\\
\noindent {\it PACS 2010: 88.05.Np; 92.30.Np; 89.60.-k; 92.60.Ry; 92.70.Aa; 92.70.Mn}
\end{abstract}
\end{frontmatter}

\section{\label{sec:Introduction}Introduction}

The Paris Agreement, available in full from \cite{ParisAgreement}, was introduced to the general public (see \cite{ParisAgreementDescription}, under ``What is the Paris Agreement?'') as follows.\\
``The Paris Agreement is a legally binding international treaty on climate change. It was adopted by $196$ Parties at the United Nations (UN) Climate Change Conference (COP21) in Paris, France, on 12 December 
2015. It entered into force on 4 November 2016. Its overarching goal is to hold `the increase in the global average temperature to well below $2^\circ$C above pre-industrial levels' and pursue efforts `to 
limit the temperature increase to $1.5^\circ$C above pre-industrial levels' \dots The Paris Agreement is a landmark in the multilateral climate change process because, for the first time, a binding agreement 
brings all nations together to combat climate change and adapt to its effects.''

In Article 2 (paragraph 1) of the document, one reads: ``This Agreement \dots aims to strengthen the global response to the threat of climate change, in the context of sustainable development and efforts to 
eradicate poverty, including by:\\
(a) Holding the increase in the global average temperature to well below $2^\circ$C above pre-industrial levels and pursuing efforts to limit the temperature increase to $1.5^\circ$C above pre-industrial 
levels, recognizing that this would significantly reduce the risks and impacts of climate change \dots'' Reference to the pre-industrial temperature levels was thus made in 2015, yet the definition of the 
term `pre-industrial' is nowhere to be found in the Paris Agreement.

The first clarification of what could be understood as pre-industrial may be found in the seminal report by the Intergovernmental Panel on Climate Change (IPCC, a body of the UN) three years after the Paris 
Agreement \cite{IPCC2018}. Chapter 2 of that report ``assesses the literature on mitigation pathways that limit or return global mean warming to $1.5^\circ$C (relative to the pre-industrial base period 
1850-1900).'' Chapter 3, entitled ``Impacts of $1.5^\circ$C Global Warming on Natural and Human Systems'' (see pp.~175-311), is an accounting of the risks which our crossing of the $1.5^\circ$C threshold 
would entail, namely the ``far more severe climate change impacts, including more frequent and severe droughts, heatwaves and rainfall'' \cite{ParisAgreementDescription}, i.e., phenomena which are already 
upon us \emph{en masse} (and which, unless our current attitude - on a number of issues: consumption of fossil fuels, deforestation, pollution, overconsumption, overpopulation, and so on - towards the 
environment is readily reassessed and environmentally friendly practices are promptly employed, are bound to become ``more frequent and severe'' in the future~\footnote{I recently read the following passage 
in a UN report (see Ref.~\cite{UN23}, p.~38): ``With a climate cataclysm looming, the pace and scale of current climate action plans are wholly insufficient to effectively tackle climate change. Increasingly 
frequent and intense extreme weather events are already impacting every region on Earth. Rising temperatures will escalate these hazards further, posing grave risks.''}). Expressed in 
\cite{ParisAgreementDescription} is the thesis that ``to limit global warming to $1.5^\circ$C, greenhouse gas emissions must peak before 2025 at the latest and decline $43$~\% by 2030.'' A mere six years 
after the Paris Agreement, these goals seemed very ambitious \cite{IPCC2021,IEA2022,IPCC2023}.

Two questions remain: Is the currently observed global warming likely to be the result of anthropogenic climate impact? How likely is such a hypothesis? Considering the countless warming and cooling phases, 
which the planet has gone through for the last one billion years \cite{ClimateHistory}, these are valid questions. This subject was addressed in Ref.~\cite{IPCC2013}, wherein the following conclusions 
were drawn (see Chapter 10, p.~869):
\begin{itemize}
\item ``More than half of the observed increase in global mean surface temperature from 1951 to 2010 is very likely ($>90$~\% confidence level) due to the observed anthropogenic increase in greenhouse 
gas~\footnote{The standard list of greenhouse gases include CO$_2$, CH$_4$, N$_2$O, O$_3$, chlorofluorocarbons, and other gases associated with human activities.} concentrations.''
\item ``It is extremely likely ($>95$~\% confidence level) that human activities caused more than half of the observed increase in global mean surface temperature from 1951 to 2010.''
\item ``Greenhouse gases contributed a global mean surface warming likely ($>66$~\% confidence level) to be between $0.5^\circ$C and $1.3^\circ$C over the period 1951-2010, with the contributions from other 
anthropogenic forcings likely to be between $-0.6^\circ$C and $0.1^\circ$C, from natural forcings likely to be between $-0.1^\circ$C and $0.1^\circ$C, and from internal variability likely to be between 
$-0.1^\circ$C and $0.1^\circ$C.''
\item ``It is virtually certain ($>99$~\% confidence level) that internal variability alone cannot account for the observed global warming since 1951.''
\end{itemize}
Similar results were reported in Ref.~\cite{USGCRP2018}, see also Fig.~2.1 therein.

Those who have watched Al Gore's 2006 documentary film `An Inconvenient Truth' will surely remember the correlation between the total amount of CO$_2$ in the atmosphere and the global average absolute 
temperature, which the $45$-th Vice President of the United States was attempting to establish in a presentation, parts of which had been recorded for the purposes of the film. I was recently wondering 
whether more recent data would make Gore's arguments (which I had found compelling one-and-a-half decades ago) even more convincing. This is how I came to the temporal dependence of the CO$_2$ concentration 
(the so-called Keeling curve), obtained at the Mauna Loa Observatory, Hawaii \cite{MaunaLoa2023}: visual inspection of that plot leaves little doubt that the CO$_2$ concentration, which has already exceeded 
the level of $420$ parts per million (ppm), shows no intention to ``peak before 2025 at the latest,'' let alone any tendency towards a decrease (by any amount) by 2030, both being (according to 
\cite{ParisAgreementDescription}) prerequisites for the containment of the global average absolute temperature within $1.5^\circ$C of the pre-industrial norms. One additional plot, essentially showing the 
Keeling curve during the last $800\,000$ years \cite{RiseOfCarbonDioxide}, is both revealing and alarming. The title of the plot is: `The relentless rise of carbon dioxide'; also shown in that figure is 
another threshold, set at the CO$_2$ concentration of about $300$ ppm and bearing the text: ``For millennia, atmospheric carbon dioxide had never been above this line.''

After this long introduction, I shall address a few issues in regard to the content and structure of this technical note. In comparison with many of the available studies, I believe that there are three 
improvements: first, the uncertainties of the input datapoints are taken into account in all parts of the analysis, i.e., in the evaluation of the reference temperature, in the transformation of the data so 
that they refer to a common temporal window, in the smoothing of the altered data, and in the extraction of predictions; second, all input data are corrected so that they refer to the pre-industrial base 
period 1850-1900 \cite{IPCC2018}; third, the seasonality (month-dependent offsets) is removed from the data by subtracting from all monthly input values the pre-industrial temperature level of the 
corresponding month (separately determined for each input dataset).

The datasets, which will be analysed in this work, are described in Section \ref{sec:Data}. The methods, which will be used in the data analysis, are detailed in Section \ref{sec:Methods}. The results of 
the analysis and the conclusions drawn thereof will be given in Sections \ref{sec:Results} and \ref{sec:Conclusions}, respectively. All figures will be placed at the end of the main part of the study. The 
significance of the $1.5^\circ$C threshold above the pre-industrial temperature levels will be addressed in Appendix \ref{App:AppA}.

\section{\label{sec:Data}The datasets}

The input comprises global average temperature anomalies (GATAs), i.e., estimates for the \emph{differences} between the values of the monthly global average \emph{absolute} temperature and a reference 
(baseline) temperature level, which is obtained on the basis of subjective criteria~\footnote{The arbitrariness of the choice of the baseline temperature level may purposely be exploited, in order to advance 
interests of individuals, of non-governmental organisations, of companies, and of corporations. For the sake of example, given that the baseline temperature level and the temperature anomaly are 
\emph{anti}correlated, a baseline temperature level, obtained (after averaging temperature data) from the recent past, will inevitably result in smaller values of the temperature anomaly, i.e., values which 
are less likely to `raise an alarm', hence trigger more urgent and decisive countermeasures.} and which is meant to approximate the global average absolute temperature in the absence (or near absence) of 
(sizeable) anthropogenic environmental impact. The temporal window, which is relevant to the determination of the baseline temperature level, is routinely placed within a few centuries (at most) of the 
current era, i.e., at times when the anthropogenic environmental impact was surely sizeable in comparison with the antiquity, yet `small' in comparison with the present-day ecological `footprint' of 
humankind. The so-called pre-industrial phase is frequently employed when setting the baseline temperature level, though (as one might expect) the adjective `pre-industrial' does not have the same meaning 
for everyone: although, as aforementioned, the second half of the $19$-th century is recommended by the IPCC \cite{IPCC2018}, earlier temporal windows have been suggested as more suitable for the 
determination of the baseline temperature, e.g., see Ref.~\cite{Hawkins2017}.

All four datasets of this work have been obtained relative to baseline temperature levels involving more recent temporal windows than the one recommended by the IPCC. To enable the analysis of the datasets 
on an equal footing, the temperature anomalies had to be altered, so that they all refer to baseline temperature levels obtained from the $50$-year period between 1 January 1850 and 31 December 1899. For 
one of the datasets, records are available only after January 1880; in that case, the (month-dependent) baseline temperature levels were obtained by averaging all available data (in that dataset) before 
1900. In retrospect, I must admit that I find the use of different temporal windows in the definition of the baseline temperature levels needless, meaningless, and error-prone. I cannot come up with a 
single reason in support of such plurality, and shall promptly follow the recommendation by the IPCC \cite{IPCC2018,IPCC2023} in this work (as well as in any follow-up to this subject in the future).

My efforts notwithstanding, I have not been able to get hold of any global average \emph{absolute} temperatures. In relation to two of the solutions (NASA GISTEMP and NOAA), the global average absolute 
temperatures are simply unavailable (see \cite{Reasons} for the NASA GISTEMP dataset discussed in Section \ref{sec:GISTEMP}, containing two links sent to me by Reto A.~Ruedy). Regarding the HadCRUT5 solution, 
the extraction of global average absolute temperatures could be possible on the basis of an unspecified charge. My two communications to Berkeley Earth (the first one relating to their data processing, the 
second to the availability of original global average absolute temperatures) remained unanswered. Consequently, the results of this work represent the best I could possibly do on the basis of the information 
which is available in the public domain.

Last but not least, only `global' (land-and-ocean) data will be analysed herein, imported (into this work from their corresponding sources) on 21 October 2023, shortly after 07:00 UTC (earlier data had been 
used during the development phase of the study). For the NASA GISTEMP dataset, records are available from January 1880 on; for the remaining three datasets, since January 1850. The last entries for the 
Berkeley Earth and HadCRUT5 datasets correspond to August 2023; for the remaining two datasets, the September 2023 values are also available. The mean seasonality (bulk or residual) was removed from all 
entries, separately for each dataset.

\subsection{\label{sec:Berkeley}The Berkeley Earth dataset}

The first set of GATAs (including uncertainties for each input datapoint) \cite{Rohde2020} originates from Berkeley Earth, an independent non-profit organisation, which aims to provide high-quality scientific 
data (and analysis) relevant to environmental issues, with emphasis currently being placed on global warming. The set of monthly GATAs is available from \cite{BerkeleyEarth}, see Section `Global Temperature 
Data' therein (and expand `Global Time Series Data'). The file `Global Monthly Averages (1850 - Recent)' contains two sets of values, representing different methods of treatment of the locations of the 
Earth's surface which are covered by sea ice: the first set is obtained via an extrapolation from land-surface air temperature data (method A), whereas the second involves an extrapolation from sea-surface 
water temperature data (method B). The Berkeley Earth group recommend the use of the results with method A \cite{BerkeleyEarth}: ``We believe that the use of air temperatures above sea ice provides a more 
natural means of describing changes in Earth's surface temperature.'' The set of GATAs, obtained with method A, will be analysed in this work.

In the description of the dataset, one finds: ``As Earth's land is not distributed symmetrically about the equator, there exists a mean seasonality to the global average temperature.'' This `mean seasonality' 
(corresponding to their baseline period, from January 1951 to December 1980) is routinely removed from all reported values \cite{BerkeleyEarth}.

Reference \cite{BerkeleyEarth} also provides a list of the global average absolute temperatures (twelve values, each representing one month of the year) which are subtracted from their original data in order 
to yield the reported GATAs: these values were used in the creation of new baseline temperature levels, corresponding to pre-industrial times as recommended by the IPCC. The twelve monthly offsets in the 
global average absolute temperatures were thus redefined, to correspond to baseline temperature levels ranging from 1 January 1850 to 31 December 1899, see Table \ref{tab:BerkeleyAbsoluteTemperaturePerMonth}.

\vspace{0.5cm}
\begin{table}[h!]
{\bf \caption{\label{tab:BerkeleyAbsoluteTemperaturePerMonth}}}Mean (bulk) seasonality of the global average absolute temperatures for method A of the Berkeley Earth dataset \cite{BerkeleyEarth}, corresponding 
to the $50$-year period from 1 January 1850 to 31 December 1899. The Berkeley Earth GATAs, analysed in this work, (should) represent differences between the global average absolute temperature for each given 
month and year, and the corresponding offset of this table.
\vspace{0.25cm}
\begin{center}
\begin{tabular}{|l|c|}
\hline
Month & Global average absolute\\
 & temperature 1850-1899 ($^\circ$C)\\
\hline
\hline
January & $11.868(36)$\\
February & $12.062(31)$\\
March & $12.711(30)$\\
April & $13.651(26)$\\
May & $14.637(21)$\\
June & $15.358(19)$\\
July & $15.697(20)$\\
August & $15.502(21)$\\
September & $14.847(21)$\\
October & $13.981(22)$\\
November & $12.865(29)$\\
December & $12.173(23)$\\
\hline
\end{tabular}
\end{center}
\vspace{0.5cm}
\end{table}

\subsection{\label{sec:HadCRUT5}The HadCRUT5 dataset}

The HadCRUT5 dataset is available from \cite{HadCRUT5}; stated therein is that the ``time series are presented as GATAs (deg C) relative to 1961-1990.'' The GATAs are accompanied by confidence intervals 
associated with the $95$~\% confidence level. For the purposes of this work, $1 \sigma$ uncertainties were obtained from the values of the original datafile. The HadCRUT5 GATAs, analysed in this work, were 
obtained after subtracting the offsets (residual seasonality) of Table \ref{tab:HadCRUT5Offsets} from the original data. Additional details on this dataset may be obtained from Ref.~\cite{Morice2021}.

\vspace{0.5cm}
\begin{table}[h!]
{\bf \caption{\label{tab:HadCRUT5Offsets}}}Offsets (residual seasonality) relevant to the processing of the HadCRUT5 data \cite{HadCRUT5}, corresponding to the $50$-year period from January 1850 to December 
1899. These values were obtained from the original time series \cite{HadCRUT5}. The HadCRUT5 GATAs were corrected via the application of the offsets of this table, so that data relative to the pre-industrial 
temperature levels of this work be obtained.
\vspace{0.25cm}
\begin{center}
\begin{tabular}{|l|c|}
\hline
Month & Offset 1850-1899 ($^\circ$C)\\
\hline
\hline
January & $-0.436(32)$\\
February & $-0.424(29)$\\
March & $-0.440(29)$\\
April & $-0.392(24)$\\
May & $-0.380(21)$\\
June & $-0.333(20)$\\
July & $-0.271(19)$\\
August & $-0.262(20)$\\
September & $-0.288(19)$\\
October & $-0.322(19)$\\
November & $-0.402(22)$\\
December & $-0.421(24)$\\
\hline
\end{tabular}
\end{center}
\vspace{0.5cm}
\end{table}

\subsection{\label{sec:GISTEMP}The NASA GISTEMP dataset}

The dataset from NASA's Goddard Institute for Space Studies, Surface Temperature (GISTEMP), is available from \cite{GISTEMP}; mentioned therein is that their data represent deviations from the corresponding 
1951-1980 means. The original dataset contains no uncertainties. However, the issue of uncertainties was addressed in Ref.~\cite{Lenssen2019}, wherein it is mentioned that the ``$95$~\% uncertainties 
are near $0.05^\circ$C in the global \emph{annual} mean for the last $50$ years and increase going back further in time reaching $0.15^\circ$C in 1880.'' To enable the analysis of these data, a linear model 
of the monthly ($1 \sigma$) uncertainty was assumed between January 1880 and December 1969, whereas a constant uncertainty was assigned to all input values from January 1970 on. The NASA GISTEMP GATAs, 
analysed in this work, were obtained by subtracting the offsets (residual seasonality) of Table \ref{tab:GISTEMPOffsets} from the original data.

\vspace{0.5cm}
\begin{table}[h!]
{\bf \caption{\label{tab:GISTEMPOffsets}}}Offsets (residual seasonality) relevant to the NASA GISTEMP data (GISS Surface Temperature Analysis version 4) \cite{GISTEMP}, extracted from the $20$-year period 
from January 1880 to December 1899. These values were obtained from the original time series \cite{GISTEMP}. The NASA GISTEMP GATAs were corrected via the application of the offsets of this table, so that 
the original time series be presented as GATAs relative to the 1880-1899 temperature level.
\vspace{0.25cm}
\begin{center}
\begin{tabular}{|l|c|}
\hline
Month & Offset 1880-1899 ($^\circ$C)\\
\hline
\hline
January & $-0.313(53)$\\
February & $-0.286(46)$\\
March & $-0.257(37)$\\
April & $-0.239(32)$\\
May & $-0.224(29)$\\
June & $-0.236(22)$\\
July & $-0.176(23)$\\
August & $-0.204(24)$\\
September & $-0.190(19)$\\
October & $-0.187(25)$\\
November & $-0.227(33)$\\
December & $-0.216(25)$\\
\hline
\end{tabular}
\end{center}
\vspace{0.5cm}
\end{table}

\subsection{\label{sec:NOAA}The NOAA NCEI dataset}

The last dataset of this work comes from the National Centers for Environmental Information (NCEI), an office of the National Oceanic and Atmospheric Administration (NOAA), which is a scientific and 
regulatory agency within the United States Department of Commerce; the dataset was downloaded from \cite{NCEI}. Monthly GATAs (without uncertainties) are available therein, with respect to the $20$-th 
century baseline temperature level, i.e., involving the temporal window from January 1901 to December 2000. (I would have found it surprising by now, if the new dataset referred to any of the aforementioned 
temporal windows.) To be able to analyse the data, uncertainties were assigned in accordance with the linear model for the monthly ($1 \sigma$) uncertainties of Section \ref{sec:GISTEMP}. The monthly GATAs 
were redefined on the basis of the pre-industrial temperature levels of this work, by subtracting the offsets (residual seasonality) of Table \ref{tab:NCEIOffsets} from all values. Additional details on the 
processing yielding the original data \cite{NCEI} may be obtained from Ref.~\cite{Vose2021}.

\vspace{0.5cm}
\begin{table}[h!]
{\bf \caption{\label{tab:NCEIOffsets}}}Offsets (residual seasonality) relevant to the NOAA NCEI data, corresponding to the $50$-year period from January 1850 to December 1899. These values were obtained from 
the original time series \cite{NCEI}. The temperature anomalies were corrected via the application of the offsets of this table, so that the original time series be altered into GATAs relative to the 
pre-industrial temperature levels of this work.
\vspace{0.25cm}
\begin{center}
\begin{tabular}{|l|c|}
\hline
Month & Offset 1850-1899 ($^\circ$C)\\
\hline
\hline
January & $-0.207(25)$\\
February & $-0.184(22)$\\
March & $-0.179(18)$\\
April & $-0.180(16)$\\
May & $-0.156(16)$\\
June & $-0.159(13)$\\
July & $-0.146(14)$\\
August & $-0.149(15)$\\
September & $-0.154(13)$\\
October & $-0.186(16)$\\
November & $-0.196(18)$\\
December & $-0.183(17)$\\
\hline
\end{tabular}
\end{center}
\vspace{0.5cm}
\end{table}

\section{\label{sec:Methods}Methods}

For the purposes of smoothing, the method of Locally Weighted Scatterplot Smoothing (also known as LOWESS) was employed \cite{Cleveland1979,Cleveland1988}. A quadratic local polynomial was used, along with 
a tri-cube weight function (closeness weights). Regarding the smoothing parameter, a number of values were tried: $1-10$ and $20$~\%. This quantity regulates the smoothness of the filtered output: very small 
values yield `noisy' output, very large ones result in `oversmoothing', hence in loss of information. The results of this work were obtained with the value of $5$~\% for the smoothing parameter, a choice 
which seems to be a reasonable compromise between `noisy' and `oversmoothed' outputs (for the data analysed herein). In this work, it was not deemed important to further fine-tune the LOWESS parameters, 
pursuing the embellishment of the output.

Regarding the weighted linear regression on segments of two of the datasets, all information may be obtained from the appendices of Ref.~\cite{Matsinos2023}, in particular Section A.1 therein. In the 
relevant plots, the standard errors of the \emph{fit}, not those of new predictions, are shown: the interest herein lies in obtaining the `bulk tendency' of the input data. It will become evident from 
Figs.~\ref{fig:TemperatureAnomalyBerkeleyEarth}-\ref{fig:TemperatureAnomalyNOAANCEI} that the \emph{monthly} GATAs have already broken the $1.5^\circ$C threshold several times during the last decade.

\section{\label{sec:Results}Results}

\begin{figure}
\begin{center}
\includegraphics [width=15.5cm] {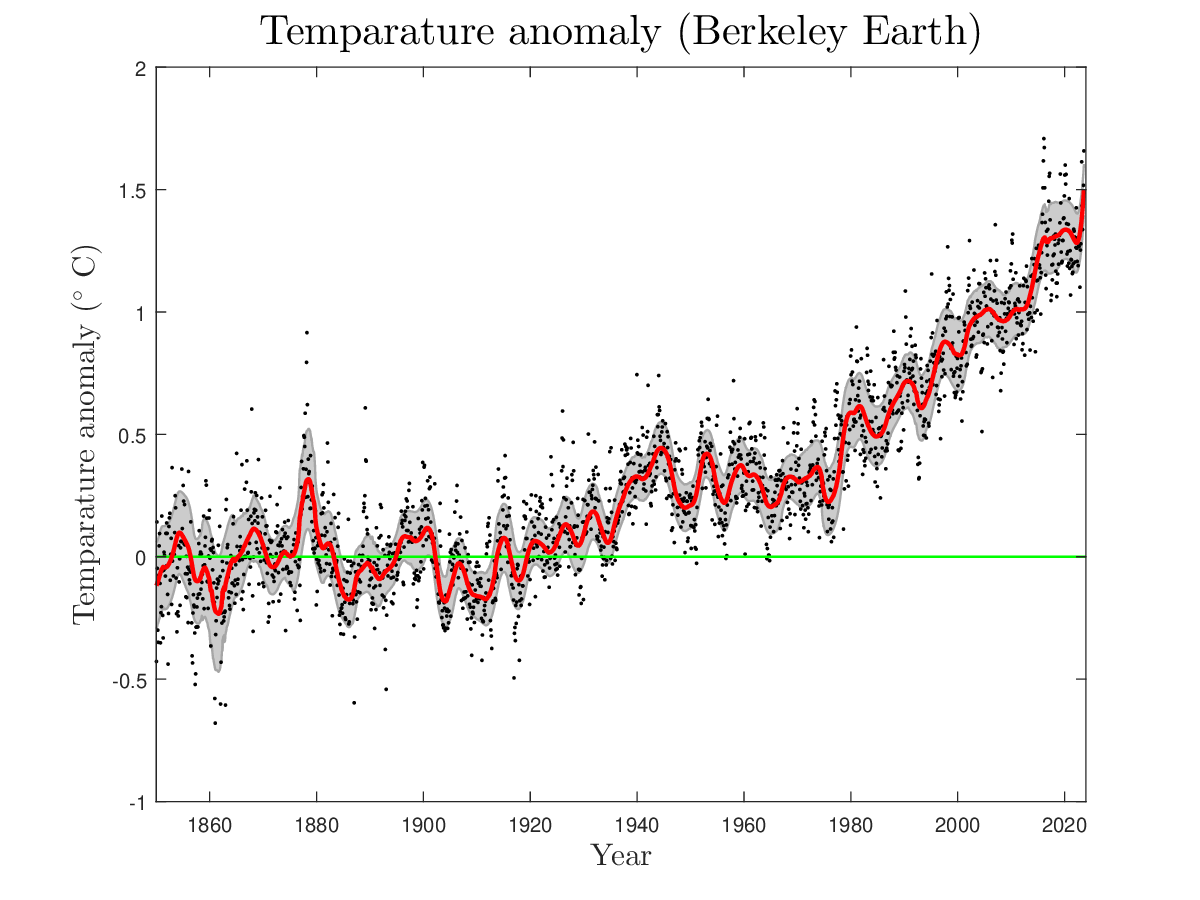}
\caption{\label{fig:TemperatureAnomalyBerkeleyEarth}The results for the Berkeley Earth GATAs, detailed in Section \ref{sec:Berkeley} and corrected on the basis of Table \ref{tab:BerkeleyAbsoluteTemperaturePerMonth} 
so that they refer to pre-1900 temperature levels. Also shown is the filtered curve (along with $1 \sigma$ uncertainties), obtained via the application of LOWESS, as explained in Section \ref{sec:Methods}. 
The uncertainties of the GATAs were used in the determination of the filtered curve, but are omitted from the figure for the sake of clarity.}
\vspace{0.5cm}
\end{center}
\end{figure}

\begin{figure}
\begin{center}
\includegraphics [width=15.5cm] {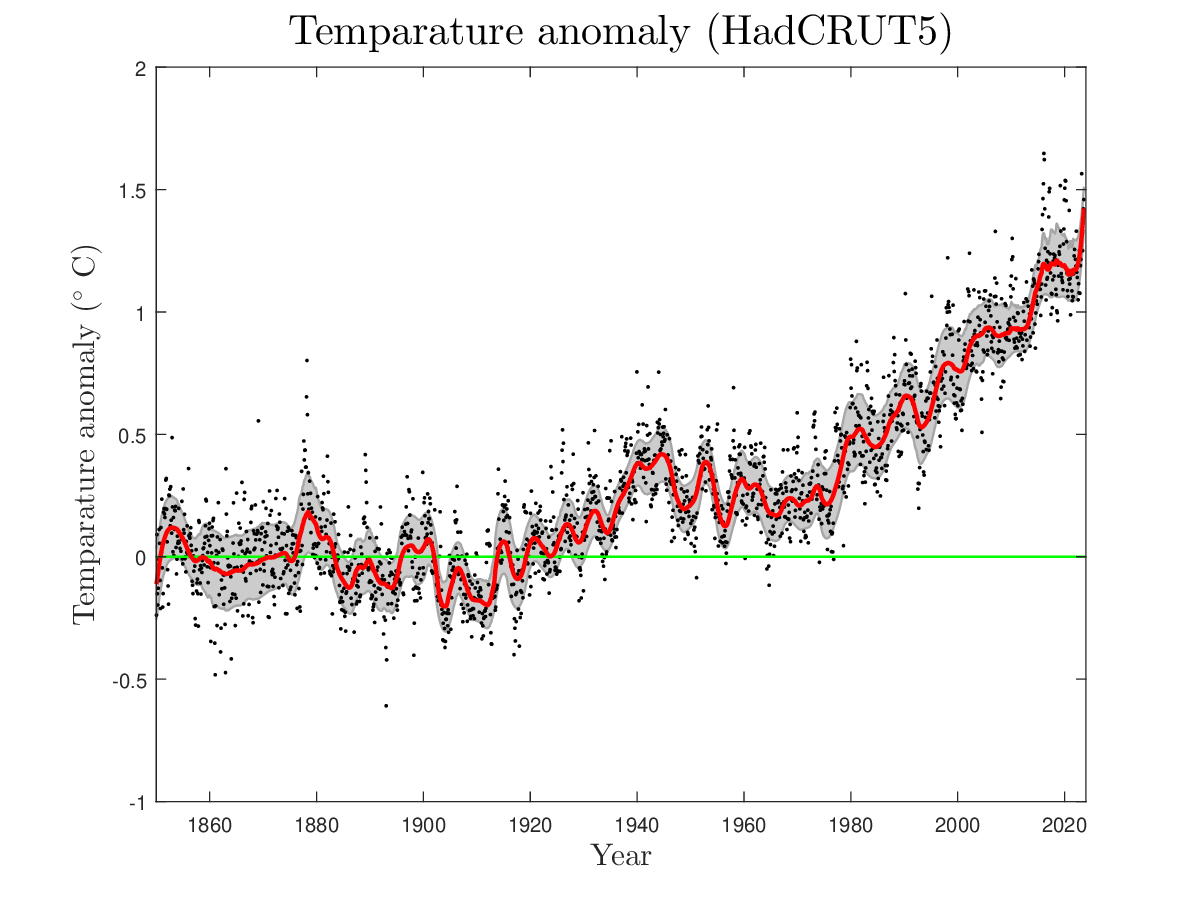}
\caption{\label{fig:TemperatureAnomalyHadCRUT5}The equivalent of Fig.~\ref{fig:TemperatureAnomalyBerkeleyEarth} for the HadCRUT5 GATAs, detailed in Section \ref{sec:HadCRUT5} and corrected on the basis 
of Table \ref{tab:HadCRUT5Offsets} so that they refer to pre-1900 temperature levels.}
\vspace{0.5cm}
\end{center}
\end{figure}

\begin{figure}
\begin{center}
\includegraphics [width=15.5cm] {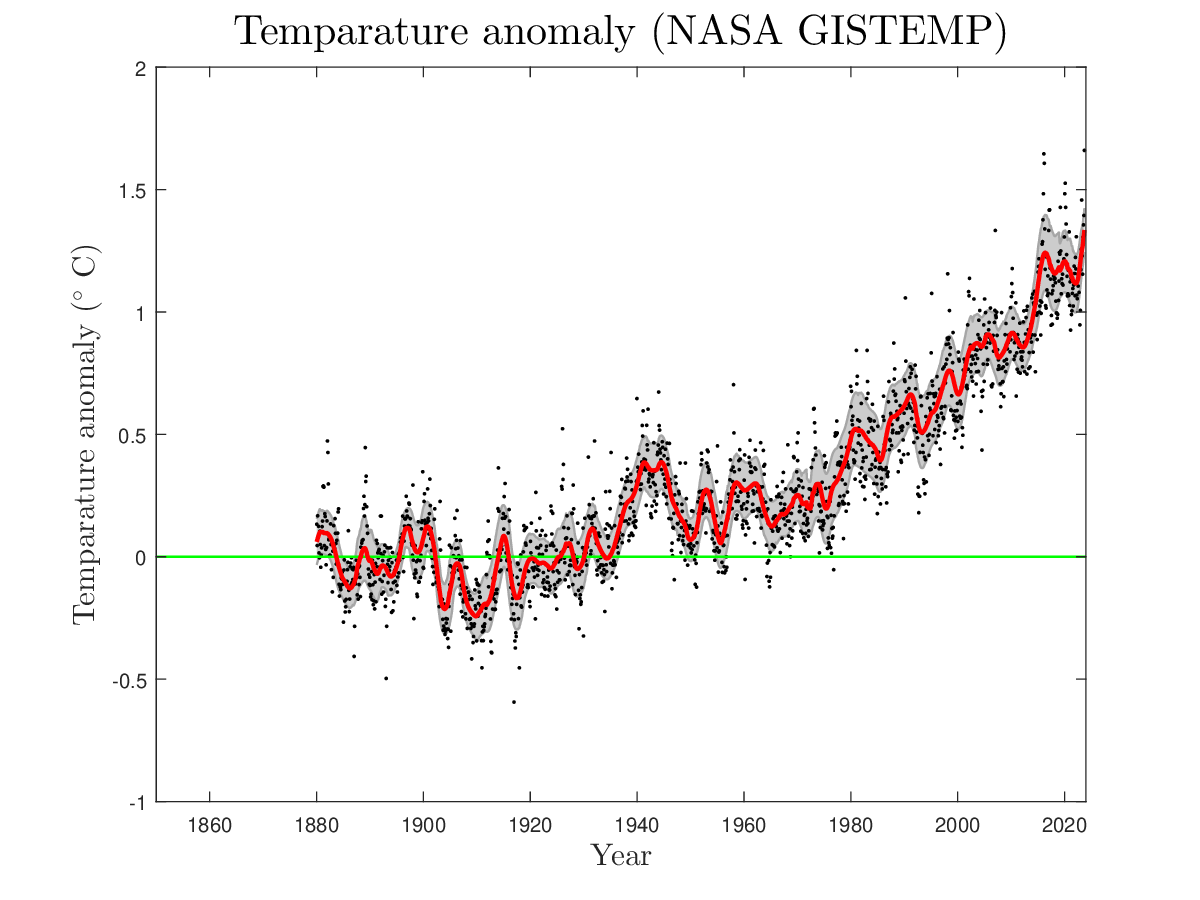}
\caption{\label{fig:TemperatureAnomalyNASAGISTEMP}The equivalent of Fig.~\ref{fig:TemperatureAnomalyBerkeleyEarth} for the NASA GISTEMP GATAs, detailed in Section \ref{sec:GISTEMP} and corrected on the 
basis of Table \ref{tab:GISTEMPOffsets} so that they refer to pre-1900 temperature levels. It must be recalled that the NASA GISTEMP GATAs are available after January 1880.}
\vspace{0.5cm}
\end{center}
\end{figure}

\begin{figure}
\begin{center}
\includegraphics [width=15.5cm] {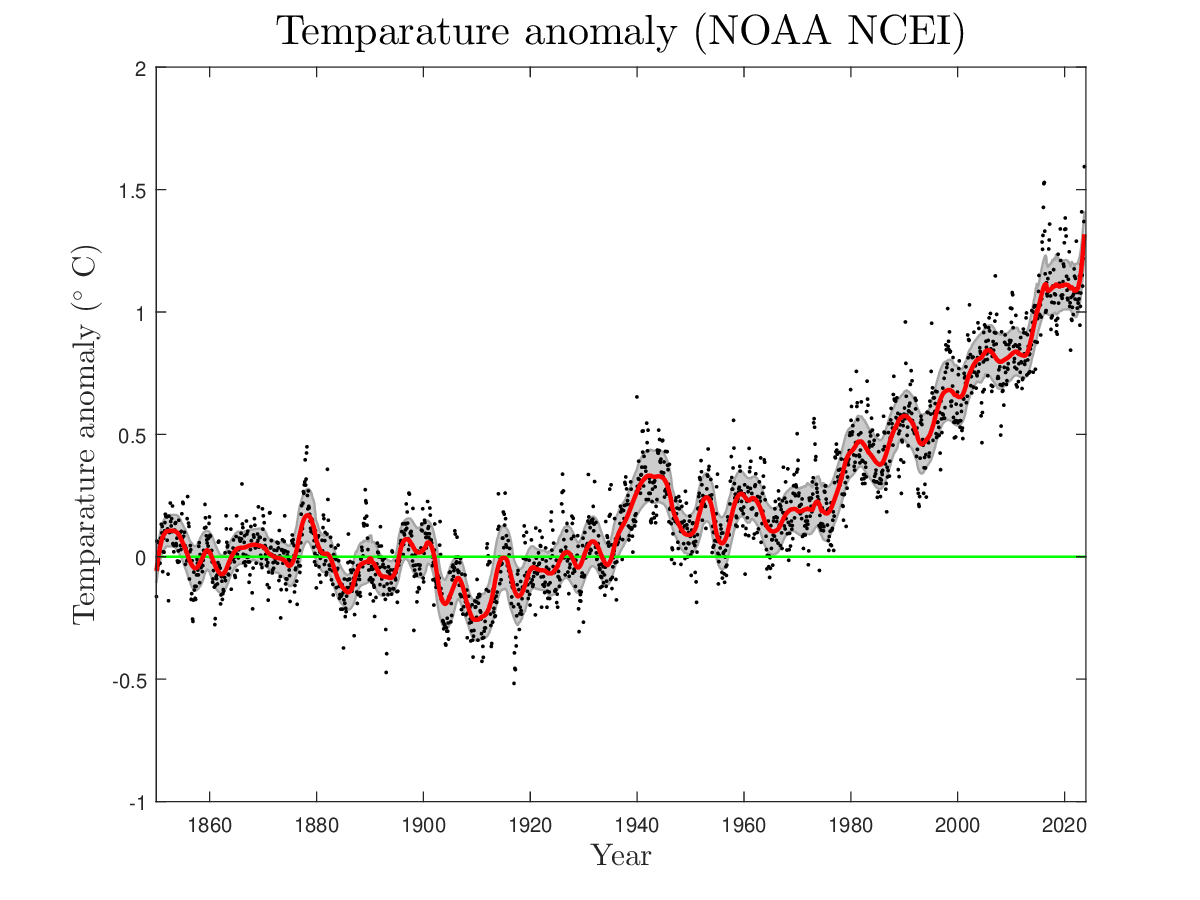}
\caption{\label{fig:TemperatureAnomalyNOAANCEI}The equivalent of Fig.~\ref{fig:TemperatureAnomalyBerkeleyEarth} for the NOAA NCEI GATAs, detailed in Section \ref{sec:NOAA} and corrected on the basis of 
Table \ref{tab:NCEIOffsets} so that they refer to pre-1900 temperature levels.}
\vspace{0.5cm}
\end{center}
\end{figure}

The results of the analysis of the four sets of GATAs, detailed in Sections \ref{sec:Berkeley}-\ref{sec:NOAA} and corrected on the basis of Tables \ref{tab:BerkeleyAbsoluteTemperaturePerMonth}-\ref{tab:NCEIOffsets} 
so that they all refer to pre-1900 temperature levels, are shown in Figs.~\ref{fig:TemperatureAnomalyBerkeleyEarth}-\ref{fig:TemperatureAnomalyNOAANCEI}. Also shown in these figures are the filtered 
curves (and corresponding $1 \sigma$ uncertainties) obtained via the application of LOWESS, as explained in Section \ref{sec:Methods}. The uncertainties of the GATAs were used, but are omitted from the plot 
for the sake of clarity. It is noticeable that \emph{all monthly global average absolute temperatures after the late 1970s, in all four datasets, exceed the pre-industrial temperature norms}! (I am referring 
to the central values of the corrected data.)

\begin{figure}
\begin{center}
\includegraphics [width=15.5cm] {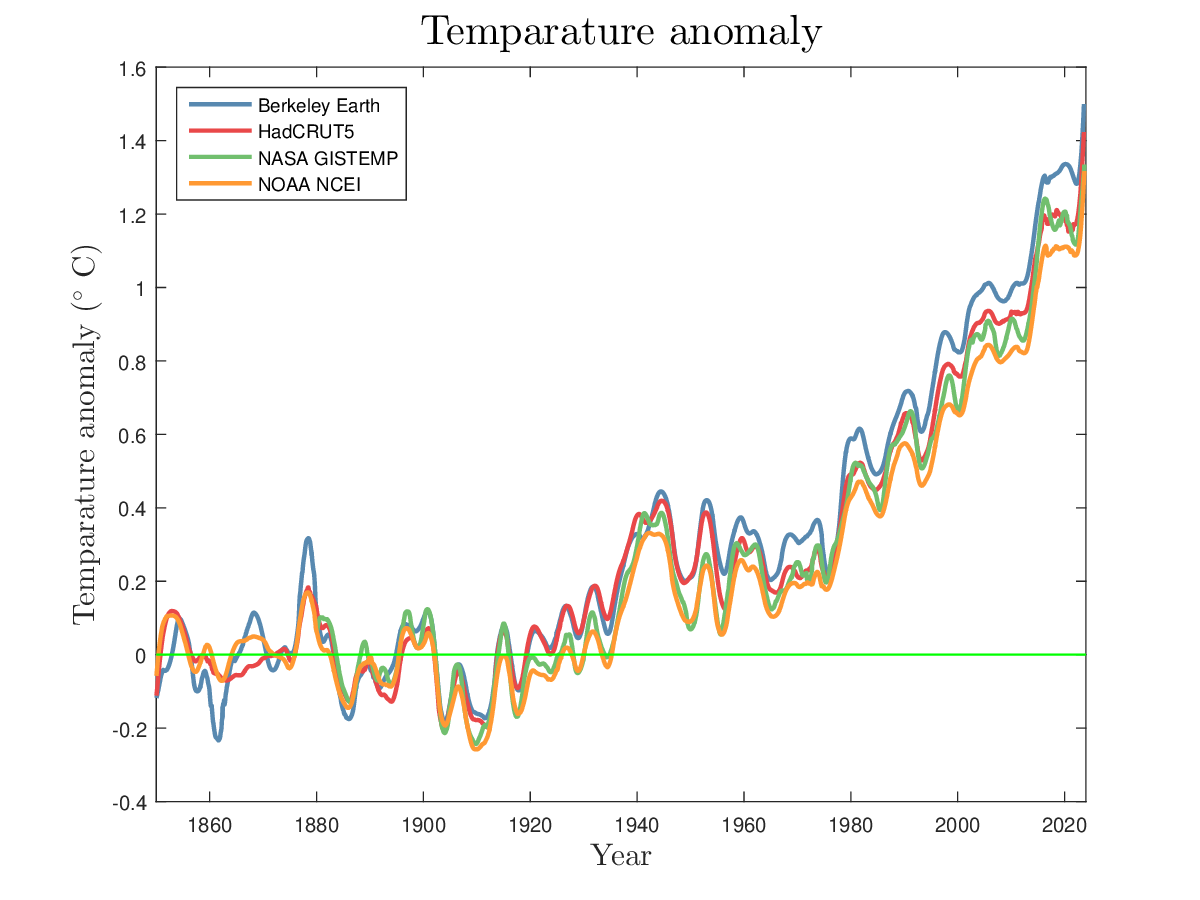}
\caption{\label{fig:TemperatureAnomaly}The filtered curves (shown without their estimated uncertainties), corresponding to the four sets of GATAs analysed in this work.}
\vspace{0.5cm}
\end{center}
\end{figure}

The four filtered curves, corresponding to these GATAs, are separately shown in Fig.~\ref{fig:TemperatureAnomaly}: for the purposes of this figure, all input datapoints are omitted, so that the four curves 
be easily compared with each other. Figure \ref{fig:TemperatureAnomaly} demonstrates that the four datasets of this work basically tell the same story: the GATAs remained close to (or below) the baseline 
temperature levels until about 1910, then gradually increased (by about $0.30^\circ$C) until about the early 1940s. A rough plateau may be seen during the subsequent three decades, after which the GATAs have 
systematically increased (by about $0.2^\circ$C per decade) until the present time.

\begin{figure}
\begin{center}
\includegraphics [width=15.5cm] {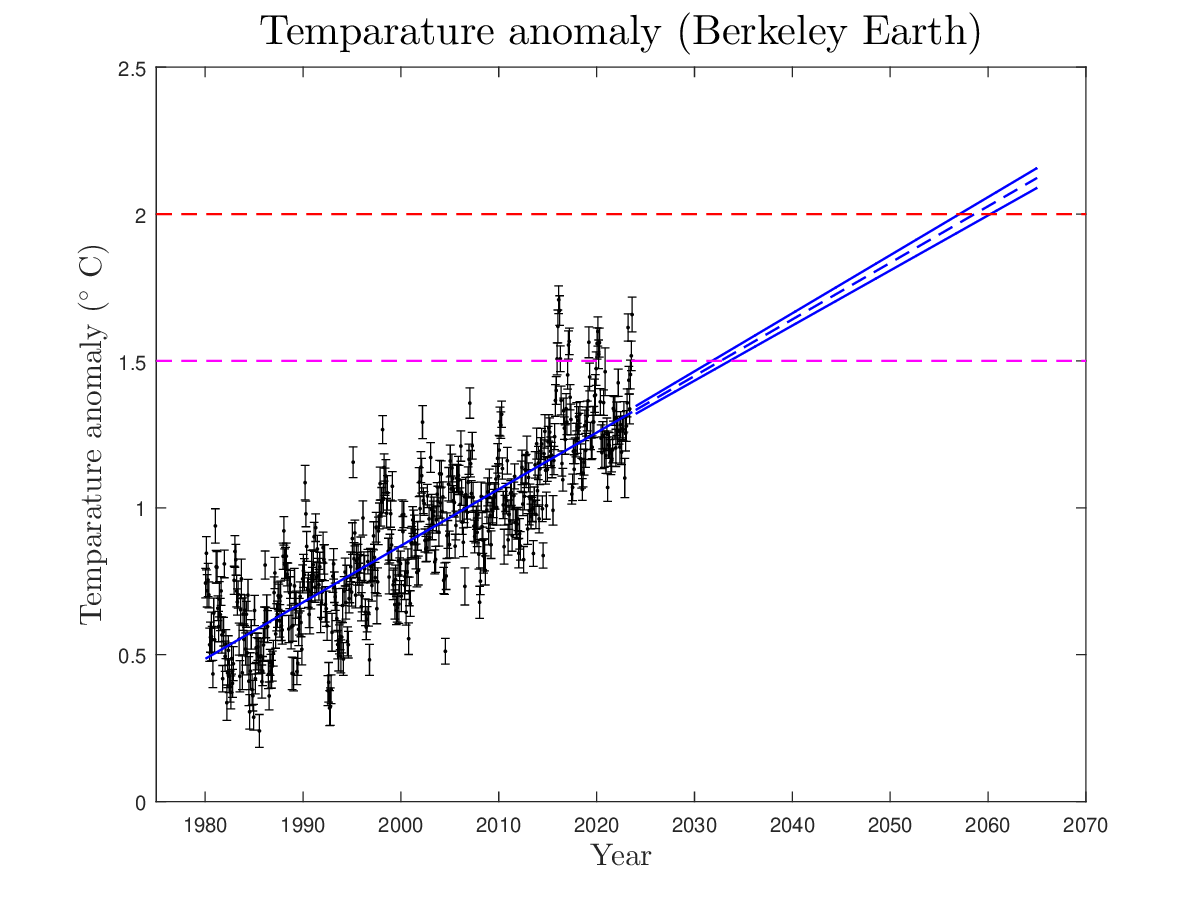}
\caption{\label{fig:TemperatureAnomalyAndPredictionsBerkeleyEarth}The results of the weighted linear regression on the Berkeley Earth GATAs (shown with their corresponding uncertainties) after 1980. The 
blue straight line (continuous up to the present time, dashed in the future) represents the optimal description of the input datapoints. The two curves around the optimal straight line, shown after the 
present time, represent predictions based on the standard errors of the fit (i.e., not on the standard errors of new predictions, see also Appendix A.1 of Ref.~\cite{Matsinos2023}). The two horizontal 
straight lines mark the $1.5^\circ$C and $2^\circ$C thresholds above the pre-industrial temperature levels.}
\vspace{0.5cm}
\end{center}
\end{figure}

\begin{figure}
\begin{center}
\includegraphics [width=15.5cm] {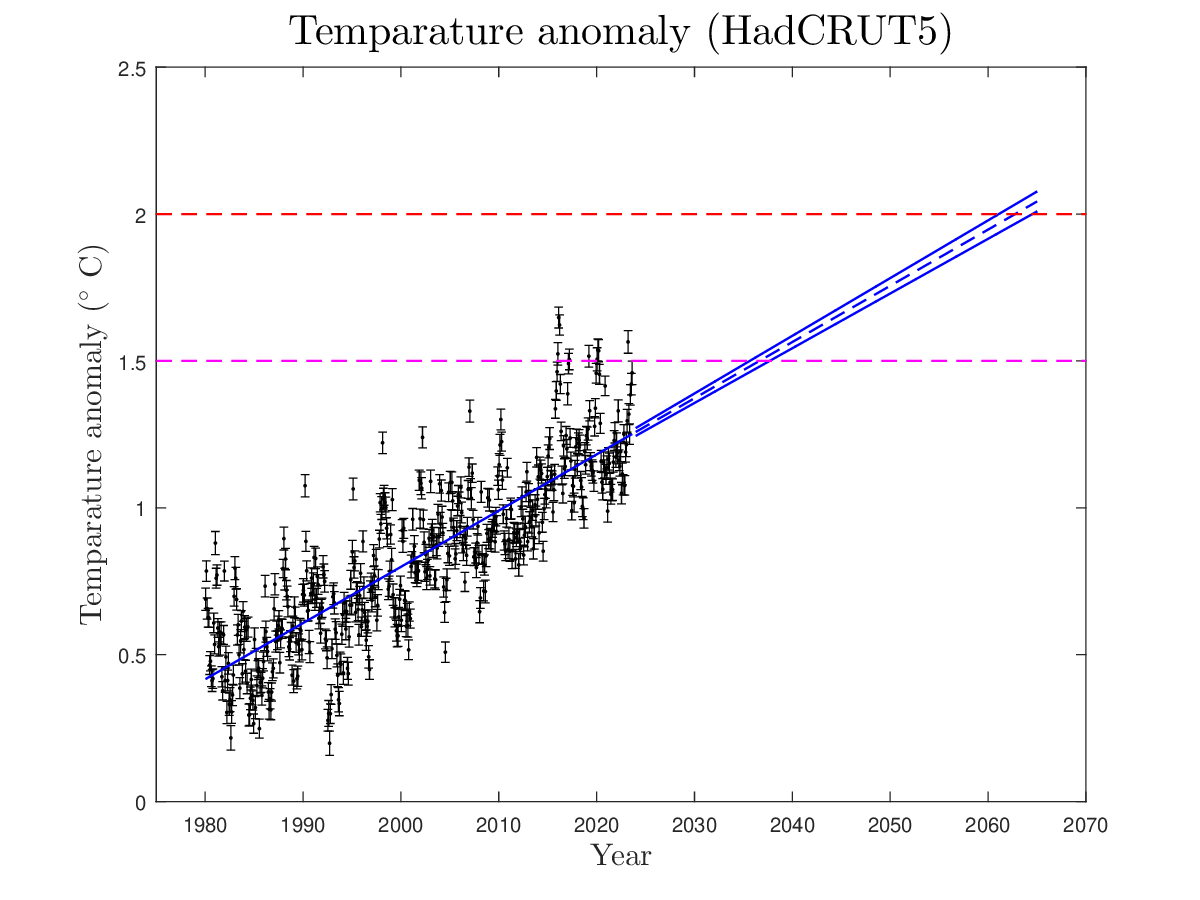}
\caption{\label{fig:TemperatureAnomalyAndPredictionsHadCRUT5}The equivalent of Fig.~\ref{fig:TemperatureAnomalyAndPredictionsBerkeleyEarth} for the HadCRUT5 dataset.}
\vspace{0.5cm}
\end{center}
\end{figure}

Visual inspection of Figs.~\ref{fig:TemperatureAnomalyBerkeleyEarth}-\ref{fig:TemperatureAnomalyNOAANCEI} suggests that the GATAs after 1980 are linearly correlated with time. The two sets of input 
datapoints, which are accompanied by meaningful (determined, not assigned) uncertainties (i.e., the Berkeley Earth and HadCRUT5 datasets), were (separately) submitted to weighted linear regression; the 
results are shown in Table \ref{tab:Optimisation}. Figures \ref{fig:TemperatureAnomalyAndPredictionsBerkeleyEarth} and \ref{fig:TemperatureAnomalyAndPredictionsHadCRUT5} show the input datapoints (with 
their corresponding uncertainties), the fitted straight lines, and predictions involving the standard errors of the fit (as explained at the end of Section \ref{sec:Methods}). These two datasets suggest 
that the $1.5^\circ$C threshold above pre-industrial temperature levels will be crossed earlier than initially expected \cite{ParisAgreement,IPCC2021}: in mid-2032 in case of the Berkeley Earth dataset, 
about four years later in case of the HadCRUT5 dataset. Of course, the accuracy of these two predictions rests upon the linearity of the effect, which (undoubtedly) does not represent the worst-case 
scenario. In addition, it is assumed that immediate counteractions will not be taken or, if they are, that they will be of no effective short-term consequence.

\vspace{0.5cm}
\begin{table}[h!]
{\bf \caption{\label{tab:Optimisation}}}Pearson correlation coefficient between the two sets of GATAs (which are accompanied by meaningful uncertainties, i.e., of the Berkeley Earth and HadCRUT5 datasets) 
and time, followed by the outcome of (separate) weighted linear regressions on these data, using time as the independent variable. The large values of the Birge factor (which is equal to the square 
root of the reduced $\chi^2_{\rm min}$) are indicative of the fluctuation which is present in the two sets of GATAs. The fitted values of the slope agree well, suggesting an increase of the global average 
absolute temperature by about $0.192^\circ$C per decade after 1980.
\vspace{0.25cm}
\begin{center}
\begin{tabular}{|l|c|c|}
\hline
Quantity & Berkeley Earth & HadCRUT5\\
\hline
\hline
Pearson correlation coefficient & $0.846$ & $0.846$\\
\hline
Birge factor & $3.08$ & $4.40$\\
Intercept ($^\circ$C) & $-37.7(1.1)$ & $-37.5(1.1)$\\
Slope ($^\circ$C yr$^{-1}$) & $0.01926(53)$ & $0.01914(53)$\\
\hline
\end{tabular}
\end{center}
\vspace{0.5cm}
\end{table}

\section{\label{sec:Conclusions}Conclusions and discussion}

A method for obtaining estimates for the time at which established global average temperature anomalies, altered so that they refer to pre-1900 temperature levels, will (on average) cross the $1.5^\circ$C 
threshold above pre-industrial (as recommended by the IPCC \cite{IPCC2018}) temperature levels was presented in this study. The uncertainties of the input values were taken into account in all parts of the 
analysis, and the seasonality (month-dependent offsets) was removed from all input values.

Extracted from the two sets of input datapoints, which are accompanied by meaningful (determined, not assigned) uncertainties (i.e., from the Berkeley Earth \cite{BerkeleyEarth} and HadCRUT5 \cite{HadCRUT5} 
datasets), were expectation values corresponding to mid-2032 and mid-2036, respectively. Provided the validity of the association of the $1.5^\circ$C threshold with the emergence of non-linearity in the 
frequency and severity of extreme-weather phenomena (see Appendix \ref{App:AppA}), the results of this straightforward study are both disturbing and alarming: in regard to extreme-weather phenomena, it 
currently appears that humankind is on the brink of entering uncharted waters.

Similar results may be found in numerous reports in journals and newspapers, in press releases of various organisations, in blogs, and the like. I mention all these sources of information without any 
intention of devaluation of their contributions to the current effort, to bring the main message home to the general public, i.e., to individuals who are not expected to read the `thousands of pages' of 
detailed reports produced by the IPCC and by similar bodies. Regarding the delivery of the message to the general public, my opinion has always been that successful politicians (like Al Gore) are very 
important in the context of the subject of this work, because they have a clear advantage over the majority of us researchers: they have been trained to sieve the `complex' to its essential features, and to 
present the `crux of the matter' to the general public in a convincing manner. Given that, as a rule, researchers a) are usually lost in the technicalities of their subjects and b) are hardly used to 
explaining complex issues in a language which may easily be understood (frequently even by other scientists, let alone the general public), I have serious doubts that they can be effective in raising social 
awareness regarding the environmental sustainability, as well as in imparting the urgency of an immediate change in our collective mindset and attitude towards the environment.

Regarding other scientific works which have dealt with the subject of this study, the following remarks are due. A sizeably larger uncertainty, as to when the $1.5^\circ$C threshold will be crossed, was 
reported in Ref.~\cite{IPCC2021} in 2021.
\begin{itemize}
\item Under the title `Faster warming' in the general description of the document, one reads: ``The report provides new estimates of the chances of crossing the global warming level of $1.5^\circ$C in the 
next decades, and finds that unless there are immediate, rapid and large-scale reductions in greenhouse gas emissions, limiting warming to close to $1.5^\circ$C or even $2^\circ$C will be beyond reach.''
\item Under B.1 therein, one further reads: ``Global surface temperature will continue to increase until at least mid-century under all emissions scenarios considered. Global warming of $1.5^\circ$C and 
$2^\circ$C will be exceeded during the $21$-st century unless deep reductions in CO$_2$ and other greenhouse gas emissions occur in the coming decades.''
\end{itemize}
Copernicus, the Earth observation component of the European Union Space Programme, had predicted, also in 2021, that the crossing of the $1.5^\circ$C threshold will occur in January 2034, e.g., see figure 
in \cite{Copernicus2021}. A similar (to the one carried out in this work), though more sophisticated, analysis is presented in Ref.~\cite{Diffenbaugh2023}; regarding the crossing of the $1.5^\circ$C threshold, 
the authors comment: ``The central estimate for the $1.5^\circ$C global warming threshold is between 2033 and 2035.'' This work endorses the findings of Refs.~\cite{Copernicus2021,Diffenbaugh2023}. An even 
bleaker near future is predicted in Ref.~\cite{Jones2023}: ``The planet is on track to reach the $1.5^\circ$C average by the 2030s.''

\begin{ack}
I am indebted to Sandra Rayne (NOAA NCEI) and to Reto A.~Ruedy (NASA Goddard Institute for Space Studies) for the clarification of questions regarding the NOAA NCEI and NASA GISTEMP datasets, respectively. 
Regarding the availability of the global average absolute temperatures corresponding to the HadCRUT5 solution, I acknowledge an exchange of e-mail with an individual who simply signed as `Jennifer' (Weather 
Desk, Met Office, United Kingdom).

The figures of this paper were created with MATLAB$\textsuperscript{\textregistered}$ (The MathWorks, Inc., Natick, Massachusetts, United States).

I have no affiliations with or involvement in any organisation, institution, company, or firm with/without financial interest in the subject matter of this paper.
\end{ack}

\clearpage
\newpage

\clearpage
\newpage
\appendix
\section{\label{App:AppA}The significance of the crossing of the $1.5^\circ$C threshold}

Before entering the subject of this appendix, a few words on the meaning of non-linearity in the context of global warming are due. If a physical quantity $y$ depends linearly on the global average absolute 
temperature $T$, then the relation
\begin{equation} \label{eq:EQA001}
y - y_0 = a (T - T_0)
\end{equation}
holds, where $a$ is a constant and $y_0$ denotes the expectation value of $y$ at the pre-industrial temperature level $T_0$. The quantity $y$ may represent:
\begin{itemize}
\item the intensity of a weather phenomenon (e.g., the amount of precipitation in a location per unit time);
\item the duration of a phenomenon (e.g., the number of hours of persistent precipitation in a location);
\item the frequency of occurrence of a phenomenon (e.g., how often precipitation of more than $20$mm occurs in a location);
\item a total amount associated with a phenomenon (e.g., the total amount of uninterrupted precipitation in a location);
\item an impact on infrastructure (e.g., the damages due to extreme-weather phenomena in a location), and so on.
\end{itemize}

Provided that linearity is fulfilled, the relation between two observations $y_1$ and $y_2$ (representing effects) corresponding to two temperature levels $T_1$ and $T_2$ (representing causes) emerges from 
Eq.~(\ref{eq:EQA001}) as
\begin{equation} \label{eq:EQA002}
\frac{y_2 - y_0}{y_1 - y_0} = \frac{T_2 - T_0}{T_1 - T_0} \, \, \, .
\end{equation}
Consequently,
\begin{equation} \label{eq:EQA003}
y_2 = y_0 + \left( y_1 - y_0 \right) \frac{T_2 - T_0}{T_1 - T_0} \, \, \, .
\end{equation}
A departure from linearity between causes and effects suggests that Eq.~(\ref{eq:EQA003}) does not hold. As far as the global warming is concerned, non-linearity is usually associated with \emph{more} 
frequent and severe effects (i.e., extreme-weather phenomena) than those anticipated on the basis of Eq.~(\ref{eq:EQA003}), in which case non-linearity is established when
\begin{equation} \label{eq:EQA004}
y_2 > y_0 + \left( y_1 - y_0 \right) \frac{T_2 - T_0}{T_1 - T_0} \, \, \, .
\end{equation}
Of course, when the observations $y$ are accompanied by measurement uncertainties, Eq.~(\ref{eq:EQA004}) must be understood in the context of statistical significance. Finally, to determine the temperature 
corresponding to the earliest departure from linearity, additional observations are required.

To the best of my knowledge, the significance of the $1.5^\circ$C threshold had not been recognised before 2016: neither does this threshold emerge as of principal significance in the Paris Agreement, nor 
does it appear as more important than any other in the earlier IPCC reports. In that respect, Ref.~\cite{IPCC2018} is equally nebulous, casually mentioning non-linearity and warning (in a few places) that 
``many risks increase non-linearly with increasing global mean surface temperature,'' but providing no solid evidence that this threshold is more significant than any other.

The importance of the $1.5^\circ$C threshold seems to have been recognised (for the first time) in Ref.~\cite{Schleussner2016}, which is essentially a comparative study of the effects associated with the 
temperature increases of $1.5^\circ$C and $2^\circ$C above the pre-industrial temperature levels. However, though the ``results reveal substantial differences in impacts between a $1.5^\circ$C and $2^\circ$C 
warming,'' they do not necessarily establish a linkage between the $1.5^\circ$C threshold and the earliest occurrence of non-linearity in the simulated phenomena.

A few years later, two studies linked the $1.5^\circ$C threshold to the outset of non-linearity. Reference \cite{King2018} concludes that ``the benefits of limiting global warming to the lower Paris Agreement 
target of $1.5^\circ$C are substantial with respect to population exposure to heat, and should impel countries to strive towards greater emissions reductions.'' Reference \cite{Barcikowska2018} is also a 
comparative study of specific effects (winter storminess and extreme precipitation) associated with the $1.5^\circ$C and $2^\circ$C thresholds. Therein, a poleward shift of the midlatitude jet stream of the 
Northern Hemisphere is predicted to occur ``mainly after exceeding the $1.5^\circ$C global warming level.'' The consequences of this disruption in the flow of the jet stream are twofold: a) less precipitation 
across the Bay of Biscay and the North Sea, and b) more precipitation, wind extremes, and storminess over the ``northwestern coasts of the British Isles, Scandinavia and the Norwegian Sea, and over the North 
Atlantic east of Newfoundland.''

It must be noted that the results of the aforementioned studies were obtained from simulations involving atmospheric/climate models which (obviously) have not been validated in the domain where 
predictions are extracted; given the potential hazards, one may as well wish that they never will.

To summarise, there is moderate-to-strong indication (personal assessment) at this time that the $1.5^\circ$C threshold \emph{does} represent a turning point in regard to climate-change issues.


\begin{thebibliography}{99}

\bibitem{ParisAgreement} https://unfccc.int/files/essential\_background/convention/application/\\pdf/english\_paris\_agreement.pdf
\bibitem{ParisAgreementDescription} https://unfccc.int/process-and-meetings/the-paris-agreement
\bibitem{IPCC2018} IPCC, `Global Warming of $1.5^\circ$C', Cambridge University Press (2018). DOI: 10.1017/9781009157940
\bibitem{UN23} United Nations, `The Sustainable Development Goals Report: Special edition' (2023). ISBN-13: 978-92-1-101460-0; https://unstats.un.org/sdgs/report/2023/
\bibitem{IPCC2021} IPCC, `Climate Change 2021: The Physical Science Basis', Cambridge University Press (2021). DOI: 10.1017/9781009157896
\bibitem{IEA2022} International Energy Agency, `CO2 Emissions in 2022';\\https://iea.blob.core.windows.net/assets/3c8fa115-35c4-4474-b237-1b00424c8844/CO2Emissionsin2022.pdf
\bibitem{IPCC2023} IPCC, `Climate Change 2023: Synthesis Report' (2023). DOI: 10.59327/IPCC/AR6-9789291691647
\bibitem{ClimateHistory} https://en.wikipedia.org/wiki/List\_of\_periods\_and\_events\_in\_climate\_history
\bibitem{IPCC2013} IPCC, `Climate Change 2013: The Physical Science Basis', Cambridge University Press (2013). ISBN-13: 978-11-0-705799-1
\bibitem{USGCRP2018} U.S.~Global Change Research Program, `Impacts, Risks, and Adaptation in the United States: Fourth National Climate Assessment, Volume II' (2018). DOI: 10.7930/NCA4.2018
\bibitem{MaunaLoa2023} https://scrippsco2.ucsd.edu/
\bibitem{RiseOfCarbonDioxide} https://climate.nasa.gov/climate\_resources/24/graphic-the-relentless-rise-of-carbon-dioxide/
\bibitem{Hawkins2017} Ed.~Hawkins \etal, `Estimating changes in global temperature since the pre-industrial period', Bull.~Amer.~Meteor.~Soc.~98(9), 1841 (2017). DOI: 0.1175/BAMS-D-16-0007.1
\bibitem{Reasons} https://data.giss.nasa.gov/gistemp/faq/\#q101; https://data.giss.nasa.gov/gistemp/faq/abs\_temp.html 
\bibitem{Rohde2020} R.A.~Rohde, Z.~Hausfather, `The Berkeley Earth land/ocean temperature record', Earth Syst.~Sci.~Data 12(4), 3469 (2020). DOI: 10.5194/essd-12-3469-2020
\bibitem{BerkeleyEarth} https://berkeleyearth.org/data
\bibitem{HadCRUT5} https://www.metoffice.gov.uk/hadobs/hadcrut5/data/current/download.html
\bibitem{Morice2021} C.P.~Morice \etal, `An updated assessment of near-surface temperature change from 1850: the HadCRUT5 dataset', J.~Geophys.~Res.~Atmos.~126(3), e2019JD032361 (2021). DOI: 10.1029/2019JD032361
\bibitem{GISTEMP} GISTEMP Team, GISS Surface Temperature Analysis (GISTEMP), version 4, NASA Goddard Institute for Space Studies;\\https://data.giss.nasa.gov/gistemp/
\bibitem{Lenssen2019} N.~Lenssen \etal, `Improvements in the GISTEMP uncertainty model', J.~Geophys.~Res.~Atmos.~124(12), 6307 (2019). DOI: 10.1029/2018JD029522
\bibitem{NCEI} https://www.ncei.noaa.gov/access/monitoring/global-temperature-anomalies/anomalies
\bibitem{Vose2021} R.S.~Vose \etal, `Implementing full spatial coverage in NOAA's global temperature analysis', Geophys.~Res.~Lett.~48, e2020GL090873 (2021). DOI: 10.1029/2020GL090873
\bibitem{Cleveland1979} W.S.~Cleveland, `Robust locally weighted regression and smoothing scatterplots', J.~Am.~Stat.~Assoc.~74(368), 829 (1979). DOI: 10.2307/2286407
\bibitem{Cleveland1988} W.S.~Cleveland, S.J.~Devlin, `Locally weighted regression: An approach to regression analysis by local fitting', J.~Am.~Stat.~Assoc.~83(403), 596 (1988). DOI: 10.2307/2289282
\bibitem{Matsinos2023} E.~Matsinos, `A discussion on two celebrated examples of application of linear regression', {\tt arXiv:2310.00915 [physics.hist-ph]}.\\DOI: 10.48550/arXiv.2310.00915
\bibitem{Copernicus2021} https://climate.copernicus.eu/how-close-are-we-reaching-global-warming-15degc
\bibitem{Diffenbaugh2023} N.S.~Diffenbaugh, E.A.~Barnes, `Data-driven predictions of the time remaining until critical global warming thresholds are reached', Proc.~Natl.~Acad.~Sci.~U.S.A.~120(6), e2207183120 (2023).\\DOI: 10.1073/pnas.2207183120
\bibitem{Jones2023} N.~Jones, `When will global warming actually hit the landmark $1.5^\circ$C limit?', Nature 618, 20 (2023). DOI: 10.1038/d41586-023-01702-w
\bibitem{Schleussner2016} C.-F.~Schleussner \etal, `Differential climate impacts for policy-relevant limits to global warming: the case of $1.5^\circ$C and $2^\circ$C', Earth Syst.~Dynam.~7, 327 (2016). 
DOI: 10.5194/esd-7-327-2016; see also Corrigendum, DOI: 10.5194/esd-7-327-2016-corrigendum
\bibitem{King2018} A.D.~King \etal, `Reduced heat exposure by limiting global warming to $1.5^\circ$C', Nat.~Clim.~Change 8, 549 (2018). DOI: 10.1038/s41558-018-0191-0
\bibitem{Barcikowska2018} M.J.~Barcikowska \etal, `Euro-Atlantic winter storminess and precipitation extremes under $1.5^\circ$C vs.~$2^\circ$C warming scenarios', Earth Syst.~Dynam.~9, 679 (2018). DOI: 
10.5194/esd-9-679-2018

\end{thebibliography}
\end{document}